\documentclass[12pt,letterpaper,hyper, notoc]{JHEP3}

\usepackage[latin1]{inputenc}
\usepackage{amsmath,epsfig}
\usepackage{amssymb,amsfonts}
\usepackage{graphicx}
\newcommand{\cN}{{\cal N}}

\newcommand{\cC}{{\cal C}}

\title{Perturbative tests of  non-perturbative counting}

\preprint{}

\author{
Atish Dabholkar$^{1, 2}$ and Jo\~ao Gomes$^{2}$\\
\it $^1${Laboratoire de Physique Th\'eorique et Hautes Energies (LPTHE)\\
\it{Universit\'e Pierre et Marie Curie-Paris 6; CNRS UMR 7589}\\
\it{Tour 24-25, 5$^{\grave{e}me}$ \'etage, Boite 126, 4 Place Jussieu}\\
\it {75252 Paris Cedex 05, France}\\
\\
\it $^2$Department of Theoretical Physics\\
\it Tata Institute of Fundamental Research\\
\it Homi Bhabha Rd, Mumbai 400 005, India\\
}\\

}

\abstract{We observe that  a class of quarter-BPS dyons in ${\cal N}=4$ theories  with charge vector $(Q, P)$ and with nontrivial values of the arithmetic duality invariant $I := \gcd (Q \wedge P)$  are nonperturbative in one frame but  perturbative in another frame. This observation suggests a test of the recently computed nonperturbative partition functions for dyons with nontrivial values of the arithmetic invariant. For all values of $I$, we  show  that the  nonperturbative counting yields vanishing indexed degeneracy for this class of states everywhere in the moduli space in precise agreement with the perturbative result.}


\begin{document}

\section{Introduction}

Quarter-BPS dyons in  string compactifications with $\mathcal{N}=4$ supersymmetry in four dimensions offer a particularly tractable framework for understanding \textit{exact} quantum properties of black holes. In some models, it has recently become possible to compute  the exact indexed degeneracies of  \textit{all} duality orbits of these dyons at \textit{all} points in the moduli space. The spectrum reveals an intricate moduli dependence and a complicated structure of walls of marginal stability which is nevertheless precisely computable.
Such detailed knowledge of the microscopic spectrum has made it possible now to probe some of the finer aspects of black hole physics such as subleading corrections to the quantum Wald entropy, both perturbative \cite{LopesCardoso:2004xf, David:2006yn} and even nonperturbative\cite{Banerjee:2008ky, Sen:2009vz, Murthy:2009dq}.

In this note we consider some simple but nontrivial perturbative tests of these nonperturbative counting formulae for quarter-BPS dyons. We consider, for purposes of illustration,  the simplest model  with   $\mathcal{N}=4$ supersymmetry obtained by compactifying the heterotic string on $T^6$. The  U-duality group $G(\mathbb{Z})$ in this case is
\begin{equation}
 SL(2,\mathbb{Z})\times O(6,22;\mathbb{Z}).
\end{equation}
In the heterotic frame, the first factor corresponds to the electric-magnetic duality  and the second factor corresponds to T-duality. A dyon with an electric charge vector $Q^i$ and magnetic charge vector $P^i$ has charge vector
\begin{equation}
 \Gamma=\left[
               \begin{array}{c}
                 Q^i \\
                 P^i \\
               \end{array}
             \right],
\end{equation}
where the index $i$ transforms in the vector representation of $O(6,22;\mathbb{Z})$ and the doublet $(Q, P)$ transforms in the fundamental representation of $SL(2,\mathbb{Z})$. If the vectors $Q$ and $P$ are parallel then the BPS state  preserves one half of the supersymmetries, if not only a quarter of the supersymmetries.

A major simplification in this model results from the fact that  all inequivalent duality orbits can be completely classified\footnote{We note in passing that a classification of duality orbits  is more  subtle for models with both higher and lower supersymmetry. For example, a complete classification of orbits of the $E_{7, 7}(\mathbb{Z})$ duality group for $\cN=8$ is not yet known. In models with $\cN=2$ supersymmetry, on the other hand, the moduli space typically receives large quantum corrections and the precise form of the arithmetic duality group itself is often not known.}.
For the purposes of dyon counting, inequivalent duality orbits are labeled essentially by a single integer $I$ which is a duality invariant\cite{Dabholkar:2007vk, Banerjee:2007sr, Banerjee:2008ri} defined by
\begin{equation}\label{arithinvariant}
 I=\text{gcd}(Q\wedge P) \, .
\end{equation}
We refer to $I$  as an arithmetic duality invariant to underscore the fact that it is an invariant of the arithmetic duality group $G(\mathbb{Z})$ which cannot be expressed as an invariant of the continuous duality group $G(\mathbb{R})$.
For each $I$, one can define the matrix of T-duality invariants
\begin{equation}
\label{matrix_charge_vector}
\Lambda  =\left(
              \begin{array}{cc}
                Q^2  & Q\cdot P \\
                Q \cdot P & P^2 \\
              \end{array}
            \right) =
            \left(
              \begin{array}{cc}
                2n & l \\
                l & 2m \\
              \end{array}
            \right)
\; .
\end{equation}
Integrality of $(n, m, l)$ follows from the fact that the charge lattice is even integral.
The counting is then captured by  a partition function $Z_I(\Omega)$ that depends on the matrix
\begin{equation}\label{period}
   \Omega = \left(
              \begin{array}{cc}
                \tau & z \\
                z & \sigma \\
              \end{array}
            \right)
\end{equation}
of the three chemical potentials $(\tau, \sigma,  z)$
conjugate to the integers $(n, m,  l)$ respectively\footnote{More accurately, one has a collection of partition functions that can be obtained from $Z_I$ by the action of an element of $SL(2, \mathbb{Z})/\Gamma^0(I)$ that is determined by  arithmetic T-duality invariants \cite{Banerjee:2008ri}.}. The dyon degeneracies for a given value of $I$ are given by Fourier coefficients of of  $Z_I(\Omega)$.  The degeneracies are moduli dependent because the Fourier coefficients  depend on the choice of the Fourier contours which in turn depend on the moduli in a precise way \cite{Dabholkar:2007vk, Sen:2007vb, Cheng:2007ch}.

These counting formulae, which we review in $\S{\ref{Nonperturbative}}$, were obtained generalizing earlier work for $I=1$ \cite{Dijkgraaf:1996it,Gaiotto:2005gf,Gaiotto:2005hc, Shih:2005uc, Shih:2005he, David:2006yn}. Partition functions for  arbitrary values  $I$ were  proposed   first from \textit{macroscopic} considerations \cite{Banerjee:2008pu, Banerjee:2008pv} so that the  Wald entropy of corresponding black holes and the structure of wall-crossings is correctly reproduced.  A microscopic two-dimensional superconformal field theory was proposed for this system in \cite{Dabholkar:2008zy} from considerations of instanton moduli space in  multi KK-monopole background.

Since there are several subtleties associated with  the instanton moduli space and  especially the moduli space of multi KK-monopoles, it is desirable to have additional tests of these counting formulae, which are at the same time  independent of considerations of Wald entropy and wall crossings. One such useful \textit{microscopic} test was devised in \cite{Sen:2007ri, Dabholkar:2008tm} by considering a class of  charge configurations that can be realized  both in string theory and in nonabelian gauge theory. Comparing with the field theory counting obtained using very different methods, one obtains a successful test of the stringy counting. Moreover, since  these states have very small charges, the test is independent of considerations from  black hole physics which corresponds to the opposite limit of large charges.

In this note we devise another independent microscopic test
the string theory counting formulae. Our strategy  will be to identify some states that are nonperturbative in one frame but are perturbative in another. A similar strategy has of course been used very  successfully for  half-BPS states in the study of  various dualities. However, in $\cN=4$ gauge theory,  the quarter-BPS states are necessarily nonperturbative and can never be mapped to any perturbative states. This is because the only perturbative BPS states in gauge theory are the gauge bosons which are half-BPS. This might lead one to expect that the same is true also in string theory. Amusingly, this is not the case and some quarter-BPS states in string theory  do map to perturbative states. Moreover, these states, even though perturbative, have \textit{nontrivial} values of the arithmetic invariant. Consequently, one obtains a simple test of the microscopic counting formulae for quarter-BPS dyons for nontrivial values of $I$ by comparing it with the  perturbative counting.

\section{A class of states \label{States}}

Without loss of generality one can work at a point in the moduli space where the $T^6$ is a product $T^4 \times T^2$ because the partition functions are independent of moduli. One can further focus on  a smaller charge sector  invariant under $SO(2,2)\subset SO(6,22)$ using U-duality.  In this sector, the dyon has the following charge configuration
\begin{equation}
    \Gamma = \left[
               \begin{array}{c}
                 Q \\
                 P \\
               \end{array}
             \right] =
             \left[
               \begin{array}{cccc}
                 {\tilde n},& n ; & {\tilde w},& w \\
                 {\tilde W},& W; & {\tilde K}, & K \\
               \end{array}
             \right].
\end{equation}
In the heterotic frame, the gauge fields associated with these charges arise from the reduction of the metric and the antisymmetric B field along a $T^2\sim S\times \tilde{S}$.
The charges $n$ and $w$ represent, respectively, the momenta and winding along the circle $S$. The $K$ charge corresponds to Kaluza-Klein monopole associated with the circle $S$, and $W$ represents the charge of NS5-branes associated with the circle $S$ but wrapping $T^4 \times \tilde{S}$. The  charges with the tilde are the analogues for the $\tilde{S}$ circle.

The charge configuration of our interest is of the following form:
 \begin{equation}\label{hetcharges}
    \Gamma_n = \left[
               \begin{array}{c}
                 Q \\
                 P \\
               \end{array}
             \right] =
             \left[
               \begin{array}{cccc}
                 0,& n ; & 0,& 0 \\
                 1,& 0; & 0, & 0 \\
               \end{array}
             \right].
\end{equation}
It is evident that the matrix of duality invariants $\Lambda$ defined in \eqref{matrix_charge_vector} vanishes for these states but the arithmetic duality invariant $I$ defined in \eqref{arithinvariant} is nevertheless nontrivial and equals $n$. Moreover, since the electric charge vector is not parallel to the magnetic charge vector the state is quarter-BPS and not half-BPS.

Under six-dimensional string-string duality, the heterotic NS5-brane is mapped to Type-IIA fundamental string, and the momenta are mapped to momenta. Thus, in the Type-II frame, our state corresponds to a perturbative Type-II  fundamental string with winding number one with $n$ units of  momentum  along the $S$ circle. We now proceed to discuss the nonperturbative and perturbative counting of these states.

\section{Nonperturbative counting \label{Nonperturbative}}

Let us summarize the prescription for extracting the nonperturbative degeneracies. At a given point $\mu$ in the moduli space,  the degeneracies for $I=1$, are given by the fourier coefficients
\begin{equation}
 d_1(\Lambda)|_\mu=\int_{\mathcal{C(\mu)}} d\Omega\, \frac{e^{-\pi i \rm{Tr}(\Omega \Lambda)}}{\Phi_{10}(\Omega)} \, .\label{siegel modular form}
\end{equation}
where ${\Phi_{10}}$ is the well-known Igusa cusp form that transforms as Siegel modular form of $Sp(2, \mathbb{Z})$ with weight ten \cite{Dijkgraaf:1996it}. The precise dependence of the contour $\cC(\mu)$ on the moduli is as in \cite{Cheng:2007ch}.

For general values of $I$, we first choose a charge configuration of the form
\begin{equation}\label{our-charge}
    \Gamma = \left[
               \begin{array}{c}
                 Q=IQ_0 \\
                 P_0 \\
               \end{array}
                \right]
\end{equation}
with $ \text{gcd}(Q\wedge P)=I$.
The degeneracies $d_I$ are then given by the Fourier coefficients of a modified elliptic genus of this superconformal field theory \cite{Dabholkar:2008zy}, and can be expressed in terms of $d_1$:
\begin{equation}
 d_I(\Lambda)|_\mu=\sum_{s|I} s \, d_1(\Lambda_s)|_\mu \, ,
 \label{deg torsion>1}
\end{equation}
where we have defined
\begin{equation}
\label{matrix_charge_vector-s}
\Lambda_s  =\left(
              \begin{array}{cc}
                Q^2/s^2  & Q\cdot P/s \\
                Q \cdot P/s & P^2 \\
              \end{array}
            \right)
\; .
\end{equation}
Note that $\Lambda_s$ thus defined has integral entries for the charge configuration \eqref{our-charge}. More general charge configurations with same values of $I$ but different values of the arithmetic T-duality invariants can be first brought to this form by an S-duality transformation \cite{Banerjee:2008ri} which is an element of $SL(2,\mathbb{Z})/\Gamma^0(I)$. Then the degeneracies are defined as above to ensure S-duality invariance.

The degeneracy defined by the formulae \eqref{siegel modular form} and \eqref{deg torsion>1} evidently has a complicated dependence on moduli. We will find though in $\S{\ref{Test}}$ that for our states which have  $\Lambda =0$, the moduli dependence disappears. Moreover, the degeneracy in fact vanishes because this particular Fourier coefficient of $1/{\Phi_{10}}$ is zero. Both these facts agree with the perturbative counting which we now describe.

\section{Perturbative counting \label{Perturbative}}

Our charge configuration (\ref{hetcharges}) maps to a perturbative state in IIA compactified on $K3 \times S^1 \times \tilde S^1$ fundamental string with unit winding and $n$ units of momentum along $S^1$. The computation of the perturbative degeneracies is straightforward but not entirely trivial. In particular, we will see that the indexed degeneracy \textit{does not} vanish for half-BPS states but \textit{does} vanish for quarter-BPS states.

For this purpose, we choose the light-cone gauge in Green-Schwarz formalism, and we work in the orbifold limit of $K3\sim T^4/\mathbb{Z}_2$. The worldsheet field thus have a target manifold $\mathbb{R}^2\times T^2\times T^4/\mathbb{Z}_2$. The fields are classified according to $Spin(2)_1\times Spin(2)_2\times SU(2)_{L}\times SU(2)_{R}$ representations where the first two $Spin(2)$ factors are the tangent space rotations in $\mathbb{R}^2$ and $T^2$ whereas the  last two $SU(2)$ factors are the tangent space rotations of the un-orbifolded $T^4$. The light-cone Green-Schwarz fermions transform as
\begin{eqnarray}
 8_s &=& (+\frac{1}{2};+\frac{1}{2};2,1) \oplus (+\frac{1}{2};-\frac{1}{2};1,2) \oplus (-\frac{1}{2};+\frac{1}{2};1,2) \oplus (-\frac{1}{2};-\frac{1}{2};2,1)\\
 8_c &=& (+\frac{1}{2};-\frac{1}{2};2,1)\oplus (+\frac{1}{2};+\frac{1}{2};1,2) \oplus (-\frac{1}{2};-\frac{1}{2};1,2) \oplus (-\frac{1}{2};+\frac{1}{2};2,1)
\end{eqnarray}
  The orbifold $\mathbb{Z}_2$ acts on the $SU(2)_R$ factor. Consequently the representation $(1,2)$ is projected out and we are left with eight real fermion zero modes.
The eight bosons transform as
\begin{equation}
 8_v=(\pm1;0;1,1)\oplus (0;\pm1;1,1) \oplus (0;0;2,2).
\end{equation}

We want to compute the partition function
\begin{equation}
 Z(q,\bar{q},y)=\text{Tr}(-1)^{F_L+F_R}q^{L_0}\bar{q}^{\bar{L}_0}y^{2J},
\end{equation}
where the trace is taken over oscillator modes. The $J$ operator is the generator of the $Spin(2)_1$. Tracing over the oscillator states gives
\begin{eqnarray}\label{fullpartition}
 Z(q,\bar{q},y)&=&(y^{\frac{1}{2}}-y^{-\frac{1}{2}})^4\prod_{n\geq1,j=\pm1}
 \frac{(1-\bar{q}^ny^{j})^2(1-q^ny^{j})^2}{(1-\bar{q}^n)^2(1-\bar{q}^ny^{2j})
 (1-q^n)^2(1-q^ny^{2j})}\times \nonumber \\
&\times&\text{Tr}_{K3}(-1)^{F_L+F_R}q^{L_0}\bar{q}^{\bar{L}_0}y^{2J}.
\end{eqnarray}
The last $K3$ factor corresponds to the  fields  which transform under $SU(2)_R$ and  get twisted. There are four bosons that transform as $(0;0;2,2)$ and fermions that transform as
\begin{eqnarray}
\text{left-moving:} \quad&& \,(+\frac{1}{2};-\frac{1}{2};1,2)\oplus(-\frac{1}{2};+\frac{1}{2};1,2) \nonumber \\
\text{right-moving:} \quad&& \,(+\frac{1}{2};+\frac{1}{2};1,2)\oplus(-\frac{1}{2};-\frac{1}{2};1,2) \, .
\end{eqnarray}
Tracing over oscilator modes of these fields gives
\begin{eqnarray}
 \text{Tr}_{K3}(-1)^{F_L+F_R}q^{L_0}\bar{q}^{\bar{L}_0}y^{2J}&=& 8\left[\frac{\vartheta_2(\tau,\nu)^2\vartheta_2(\bar{\tau},\nu)^2}
 {\vartheta_2(\tau,0)^2\vartheta_2(\bar{\tau},0)^2}
 + \frac{\vartheta_3(\tau,\nu)^2\vartheta_3(\bar{\tau},\nu)^2}
 {\vartheta_3(\tau,0)^2\vartheta_3(\bar{\tau},0)^2}
 + \frac{\vartheta_4(\tau,\nu)^2\vartheta_4(\bar{\tau},\nu)^2}
 {\vartheta_4(\tau,0)^2\vartheta_4(\bar{\tau},0)^2}\right] \nonumber \\
&+&(y^{\frac{1}{2}}-y^{-\frac{1}{2}})^4(\ldots)\nonumber
\end{eqnarray}
The last term , which we denote by dots, will not be important for our computation as it contributes with additional fermion zero modes.

The  quarter-BPS  dyons break twelve supersymmetries which lead to six complex fermion zero modes. Hence the degeneracy is captured by a helicity supertrace $B_6$  which can extracted from $Z(q,\bar{q},y)$ by acting  with six $y$ derivatives before setting $y=1$ \cite{Kiritsis:1997hj, Kiritsis:1997gu}. This particular helicity supertrace has been computed in \cite{Gregori:1997hi} and found to be zero as a result of an accidental cancelation between quarter-BPS multiplets which is not a consequence of supersymmetry. We present the computation in a slightly different form below.

Note that $Z(q,\bar{q},y)$ has explicitly a factor of $(y^{1/2}-y^{-1/2})^4$, which means that we only need to take two further derivatives on the $(q,\bar{q})$ dependent piece. That is
\begin{eqnarray}
 \frac{1}{2}\frac{d^2}{dy^2}\left[\frac{Z(q,\bar{q},y)}
 {(y^{1/2}-y^{-1/2})^4}\right]|_{y=1}&=&\frac{1}{2}\frac{d^2}
 {dy^2}\left\lbrace\prod_{n\geq1,j=\pm1}\frac{(1-\bar{q}^ny^{j})^2(1-q^ny^{j})^2}
 {(1-\bar{q}^n)^2(1-\bar{q}^ny^{2j})(1-q^n)^2(1-q^ny^{2j})}\right\rbrace_{y=1}\times 24 \nonumber \\
&&+\frac{1}{2}\frac{d^2}{dy^2}
\left\lbrace\text{Tr}_{K3}(-1)^{F_L+F_R}q^{L_0}\bar{q}^{\bar{L}_0}y^{2J}\right\rbrace_{y=1} \, .
\end{eqnarray}
Because the partition function is $q\longleftrightarrow \bar{q}$ symmetric we will have quarter-BPS states by exciting either the left or right-moving sectors. We consider the states with right-movers in the ground states with arbitrary left-moving oscillations with degeneracy $d(m)$. The generating function for these degeneracies is
\begin{eqnarray}
 \sum_{m} d(m)q^m &=& 16\times \left[\sum_{s\geq 1}\sum_{n\geq1}  s(3-(-1)^s)q^{ns} -s(1+(-1)^s)q^{(n-\frac{1}{2})s}\right] \nonumber \\
&=& 16\times \sum_{s\geq 1}\sum_{n\geq1}  s(3-(-1)^s)q^{ns} -\sum_{s\geq1}\sum_{n\geq 1} 64sq^{(n-\frac{1}{2})2s} \, .
\end{eqnarray}
The level matching condition is
\begin{equation}
 L_0-\bar{L}_0=nw=I,
\end{equation}
where $w$ and $n$ are respectively the winding and momenta along the circle $S$. The BPS condition sets $\bar{L}_0=0$. Setting $L_0=I$ for our configuration yields
\begin{equation}\label{degeneracy}
 d(I)=16\left[\sum_{s|I}s(3+(-1)^{s+1})-4\sum_{(2s+1)|I}\frac{I}{2s+1}\right]
\end{equation}
This strange sum over divisors actually vanishes. To see this note that any number $I$ can be written as $I=2^N I_{\text{odd}}$, for some $N$, where $I_{\text{odd}}$ is odd. The complete set of divisors of $I$ is
\begin{equation}
 \left\lbrace 2^i\hat{s}_j \right\rbrace
\end{equation}where $i$ goes from $0$ to $N$ and $j$ runs through the divisors of $I_{\text{odd}}$. The sum (\ref{degeneracy}) simplifies to
\begin{eqnarray}
 d(I)&=&16\left[\sum_{\hat{s}}4\hat{s}+2\hat{s}\sum_{i=1}^N 2^i-4. 2^N \hat{s}\right]\\
&=&16\left[\sum_{\hat{s}}4\hat{s}+4\hat{s}(2^N-1)-4. 2^N \hat{s}\right]=0
\end{eqnarray}
The same result was found in \cite{Gregori:1997hi} as a consequence a theta identity Eq.(B.22).

Note that for $n=0$, we actually have a  half-BPS state which is dual the perturbative state also of the heterotic string. Since it breaks only  eight supersymmetries, there are only four complex fermion zero modes. Hence   we need to take only  the fourth derivative of the partition function \eqref{fullpartition} to compute the helicity supertrace $B_4$. For $B_4$, one correctly obtains a  \textit{nonzero} multiplicity which moreover equals $24$ consistent with the heterotic counting \cite{Kiritsis:1997hj}.
We thus see that vanishing of $B_6$ in our case is accidental. It is not a consequence of cancelation between bosons and fermions within a multiplet but rather of cancelation between full supermultiplets.

\section{A test \label{Test}}

We would now like to reproduce two interesting facts about the perturbative counting from the perspective of the nonperturbative counting.

First, the degeneracy of perturbative states is expected to be moduli independent. On the other hand, the nonperturbative spectrum \textit{a priori}  has a sensitive moduli dependence. The states could even decay upon crossing walls of marginal stability. For the class of states, that we have considered, however, the moduli dependence disappears. To see this, we note first that $d_1$ is S-duality covariant
\begin{equation}
 d_1(\Lambda ')|_{\mu'}=d_{1}(\Lambda)|_{\mu} \,
\end{equation}
and thus the degeneracy of a given  configuration $\Lambda$ at a given point in the moduli space $\mu$ equals the degeneracy of the dual charge configuration $\Lambda'$ at another  point $\mu'$ in the moduli space which is an image of $\mu$ under S-duality.  Now since   $\Lambda = 0$ is invariant under with S-duality, we have
\begin{equation}
 d_1(0)|_{\mu'}=d_{1}(0)|_{\mu} \,
\end{equation}
Furthermore, using the natural embedding of the $ SL(2,\mathbb{Z})$ S-duality group in $Sp(2, \mathbb{Z})$, it is known that moduli space
is divided into chambers separated by walls and one can go from any chamber to any other chamber by an S-duality transformation.
One can thus cover the entire moduli space by S-duality transformations starting from any given chamber\cite{Cheng:2007ch}. Therefore, using \eqref{deg torsion>1} we conclude that $d_I(0)|_{\mu}$ is independent of $\mu$ for any $\mu$.

Second, it is easy to check from the expansion of the explicit expression for  $1/{\Phi_{10}}$ that $d_1(0) =0$ and hence we conclude from \eqref{deg torsion>1} that
\begin{equation}
 d_I(0)=0 \, .
\end{equation}
Note that the vanishing of $d_1(0)$ itself is a consequence
of  a peculiar fact about $1/{\Phi_{10}}$ that a particular Fourier coefficient vanishes which is not true for a general Siegel form. The vanishing of $d_I(0)$  depends in addition on  the fact that $d_I$ is  expressible in terms of $d_1$ as in \eqref{deg torsion>1}. Thus, the vanishing of degeneracies for this specific charge configurations, even though simple to verify, constitutes a nontrivial test of \eqref{deg torsion>1}.
Recall that our degeneracies are actually  indexed degeneracies  and hence they can be zero or even negative.

In conclusion, both the actual degeneracy and the moduli dependence of the nonperturbative counting   is in complete agreement with the perturbative counting.

\subsection*{Acknowledgments}

It is a pleasure to thank  Sameer Murthy and Suresh Nampuri for early collaboration. We would like to thank Boris Pioline and Ashoke Sen for comments on the draft. The work of A.~D. was supported in part by the Excellence Chair of the Agence Nationale de la Recherche (ANR). The work of J.G. was supported in part by Funda\c{c}\~{a}o para a Ci\^{e}ncia e Tecnologia (FCT).

\bibliographystyle{JHEP}
\bibliography{perturbative}

\providecommand{\href}[2]{#2}\begingroup\raggedright\begin{thebibliography}{10}

\bibitem{LopesCardoso:2004xf}
G.~Lopes~Cardoso, B.~de~Wit, J.~Kappeli, and T.~Mohaupt, {\it Asymptotic
  degeneracy of dyonic {N = 4} string states and black hole entropy},  {\em
  JHEP} {\bf 12} (2004) 075,
  [\href{http://xxx.lanl.gov/abs/hep-th/0412287}{{\tt hep-th/0412287}}].

\bibitem{David:2006yn}
J.~R. David and A.~Sen, {\it {CHL} dyons and statistical entropy function from
  {D1-D5} system},  \href{http://xxx.lanl.gov/abs/hep-th/0605210}{{\tt
  hep-th/0605210}}.

\bibitem{Banerjee:2008ky}
N.~Banerjee, D.~P. Jatkar, and A.~Sen, {\it {Asymptotic Expansion of the N=4
  Dyon Degeneracy}},  {\em JHEP} {\bf 05} (2009) 121,
  [\href{http://xxx.lanl.gov/abs/0810.3472}{{\tt 0810.3472}}].

\bibitem{Sen:2009vz}
A.~Sen, {\it {Arithmetic of Quantum Entropy Function}},  {\em JHEP} {\bf 08}
  (2009) 068, [\href{http://xxx.lanl.gov/abs/0903.1477}{{\tt 0903.1477}}].

\bibitem{Murthy:2009dq}
S.~Murthy and B.~Pioline, {\it {A Farey tale for N=4 dyons}},  {\em JHEP} {\bf
  09} (2009) 022, [\href{http://xxx.lanl.gov/abs/0904.4253}{{\tt 0904.4253}}].

\bibitem{Dabholkar:2007vk}
A.~Dabholkar, D.~Gaiotto, and S.~Nampuri, {\it Comments on the spectrum of
  {CHL} dyons},  \href{http://xxx.lanl.gov/abs/hep-th/0702150}{{\tt
  hep-th/0702150}}.

\bibitem{Banerjee:2007sr}
S.~Banerjee and A.~Sen, {\it Duality orbits, dyon spectrum and gauge theory
  limit of heterotic string theory on {$T^6$}},
  \href{http://xxx.lanl.gov/abs/arXiv:0712.0043 [hep-th]}{{\tt arXiv:0712.0043
  [hep-th]}}.

\bibitem{Banerjee:2008ri}
S.~Banerjee and A.~Sen, {\it S-duality action on discrete {T}-duality
  invariants},  \href{http://xxx.lanl.gov/abs/arXiv:0801.0149 [hep-th]}{{\tt
  arXiv:0801.0149 [hep-th]}}.

\bibitem{Sen:2007vb}
A.~Sen, {\it Walls of marginal stability and dyon spectrum in {N=4}
  supersymmetric string theories},  {\em JHEP} {\bf 05} (2007) 039,
  [\href{http://xxx.lanl.gov/abs/hep-th/0702141}{{\tt hep-th/0702141}}].

\bibitem{Cheng:2007ch}
M.~C.~N. Cheng and E.~Verlinde, {\it Dying dyons don't count},
  \href{http://xxx.lanl.gov/abs/arXiv:0706.2363 [hep-th]}{{\tt arXiv:0706.2363
  [hep-th]}}.

\bibitem{Dijkgraaf:1996it}
R.~Dijkgraaf, E.~P. Verlinde, and H.~L. Verlinde, {\it Counting dyons in {N =
  4} string theory},  {\em Nucl. Phys.} {\bf B484} (1997) 543--561,
  [\href{http://xxx.lanl.gov/abs/hep-th/9607026}{{\tt hep-th/9607026}}].

\bibitem{Gaiotto:2005gf}
D.~Gaiotto, A.~Strominger, and X.~Yin, {\it New connections between 4d and 5d
  black holes},  \href{http://xxx.lanl.gov/abs/hep-th/0503217}{{\tt
  hep-th/0503217}}.

\bibitem{Gaiotto:2005hc}
D.~Gaiotto, {\it Re-recounting dyons in {N = 4} string theory},
  \href{http://xxx.lanl.gov/abs/hep-th/0506249}{{\tt hep-th/0506249}}.

\bibitem{Shih:2005uc}
D.~Shih, A.~Strominger, and X.~Yin, {\it {Recounting dyons in N = 4 string
  theory}},  \href{http://xxx.lanl.gov/abs/hep-th/0505094}{{\tt
  hep-th/0505094}}.

\bibitem{Shih:2005he}
D.~Shih and X.~Yin, {\it Exact black hole degeneracies and the topological
  string},  {\em JHEP} {\bf 04} (2006) 034,
  [\href{http://xxx.lanl.gov/abs/hep-th/0508174}{{\tt hep-th/0508174}}].

\bibitem{Banerjee:2008pu}
S.~Banerjee, A.~Sen, and Y.~K. Srivastava, {\it Partition functions of torsion
  $ > 1$ dyons in heterotic string theory on {$T^6$}},
  \href{http://xxx.lanl.gov/abs/arXiv:0802.1556 [hep-th]}{{\tt arXiv:0802.1556
  [hep-th]}}.

\bibitem{Banerjee:2008pv}
S.~Banerjee, A.~Sen, and Y.~K. Srivastava, {\it {Generalities of quarter BPS
  dyon partition function and dyons of torsion two}},
  \href{http://xxx.lanl.gov/abs/arXiv:0802.0544 [hep-th]}{{\tt arXiv:0802.0544
  [hep-th]}}.

\bibitem{Dabholkar:2008zy}
A.~Dabholkar, J.~Gomes, and S.~Murthy, {\it {Counting all dyons in N =4 string
  theory}},  \href{http://xxx.lanl.gov/abs/0803.2692}{{\tt 0803.2692}}.

\bibitem{Sen:2007ri}
A.~Sen, {\it {Three string junction and N=4 Dyon spectrum}},  {\em JHEP} {\bf
  12} (2007) 019, [\href{http://xxx.lanl.gov/abs/arXiv:0708.3715 [hep-th]}{{\tt
  arXiv:0708.3715 [hep-th]}}].

\bibitem{Dabholkar:2008tm}
A.~Dabholkar, K.~Narayan, and S.~Nampuri, {\it Degeneracy of decadent dyons},
  \href{http://xxx.lanl.gov/abs/arXiv:0802.0761 [hep-th]}{{\tt arXiv:0802.0761
  [hep-th]}}.

\bibitem{Kiritsis:1997hj}
E.~Kiritsis, {\it {Introduction to superstring theory}},
  \href{http://xxx.lanl.gov/abs/hep-th/9709062}{{\tt hep-th/9709062}}.

\bibitem{Kiritsis:1997gu}
E.~Kiritsis, {\it {Introduction to non-perturbative string theory}},
  \href{http://xxx.lanl.gov/abs/hep-th/9708130}{{\tt hep-th/9708130}}.

\bibitem{Gregori:1997hi}
A.~Gregori {\em et~al.}, {\it {R**2 corrections and non-perturbative dualities
  of N = 4 string ground states}},  {\em Nucl. Phys.} {\bf B510} (1998)
  423--476, [\href{http://xxx.lanl.gov/abs/hep-th/9708062}{{\tt
  hep-th/9708062}}].

\end{thebibliography}\endgroup

\end{document}